\documentclass[prd,aps,floats,preprintnumbers,preprint]{revtex4}

\usepackage{graphicx}
\usepackage{multirow}
 
\newcommand{\be}{\begin{equation}}
\newcommand{\ee}{\end{equation}}
\newcommand{\bea}{\begin{eqnarray}}
\newcommand{\eea}{\end{eqnarray}}

\begin{document}

\setlength{\unitlength}{1mm}

\title{Can smooth LTB models mimicking the cosmological constant for the luminosity distance also satisfy the age constraint?}

\author{Antonio Enea Romano$^{1,2,3}$}
\email{aer@phys.ntu.edu.tw}
\affiliation{
$^1$Leung Center for Cosmology and Particle Astrophysics, National Taiwan University, Taipei 10617, Taiwan, R.O.C. \\
$^2$CERN, Theory Division,CH-1211 Geneva 23, Switzerland\\
$^3$ Instituto de F\'{\i}sica, Universidad de Antioquia, Colombia
\\
}


\begin{abstract}
The central smoothness of the functions defining a LTB solution plays a crucial role in their ability to mimick the effects of the cosmological constant. So far attention has been focused on $C^{1}$ models while in this paper we approach it a more general way, investigating the implications of higher order central smoothness conditions for LTB models reproducing the luminosity distance of a $\Lambda CDM$ Universe. Our analysis is based on a low red-shift expansion, and extends previous investigations by including also the constraint coming from the age of the Universe and re-expressing the equations for the solution of the inversion problem in a manifestly dimensionless form which makes evident the freedom to accommodate any value of $H_0$ as well, correcting some wrong claims that the observed value of $H_0$ would be enough to rule out LTB models. 

Higher order smoothness conditions strongly limit the number of possible  solutions respect to the first order condition. Neither a $C^{1}$ or a $C^{i}$ LTB model can  both satisfy the age constraint and mimick the cosmological constant for the luminosity distance.  One difference is in the case in which the age constraint is not included and the bang function is zero, in which there is a unique solution for $C^1$ models but no solution for the $C^{i}$ case.
Another difference is in the case in which the age constraint is not included and the bang function is not zero, in which the solution is undetermined for both $C^1$ and $C^{i}$ models, but the latter ones have much less residual parametric freedom.

Our results imply that any  LTB model able to fit luminosity distance data and satisfy the age constraint  is either not mimicking exactly the $\Lambda CDM$ red-shift space observables theoretical predictions or it is not $C^{\infty}$ smooth.
\end{abstract}

\maketitle

\section{Introduction}
A vast range of cosmological observables  \cite{Perlmutter:1999np,Riess:1998cb,Tonry:2003zg,Knop:2003iy,Barris:2003dq,Riess:2004nr,WMAP2003,Spergel:2006hy} when interpreted under the assumption of large scale spatial homogeneity supports a dominant dark energy component, giving rise
to a positive cosmological acceleration. This value of the cosmological constant is different from the one predicted by quantum field theory  by few orders of magnitude. Given this evident inconsistency about our understanding of the Universe, a lot of attention has been devoted to alternatives to dark energy, such as modified gravity theories or large scale inhomogeneities.
These cosmological models are based on violating some of the assumptions of standard cosmology. One of these assumptions is that the Universe is well described by general relativity at any scale, and a violation of this leads to modified gravity theories, which consist of theories of gravity which agree with general relativity on small scales such as the solar system, and depart from it on larger cosmological scales. The other main assumption of the standard cosmological model is that the Universe is homogeneous on sufficiently large scales, and can be correctly described by a spatially homogeneous and isotropic metric, i.e. the FLRW metric. Since there is no conclusive evidence of such a large scale homogeneity, and the Copernican principle is more of a philosophical principle than an experimental evidence, inhomogeneous cosmological models have been studied as alternatives to dark energy. 

It has been 
proposed \cite{Nambu:2005zn,Kai:2006ws}
 that we may be at the center of an inhomogeneous isotropic universe without cosmological constant described by a Lemaitre-Tolman-Bondi (LTB)  solution of Einstein's field 
equations, where spatial averaging over one expanding and one contracting 
region is producing a positive averaged acceleration $a_D$, but it has been shown how spatial averaging can give rise to averaged quantities which are not observable \cite{Romano:2006yc}.
A method to map luminosity distance as a function of
 redshift $D_L(z)$ to LTB models has been 
 proposed \cite{Chung:2006xh,Yoo:2008su},
 showing that an inversion method can be applied successfully to 
reproduce the observed $D_L(z)$.   
Analysis of observational data in inhomogeneous models without dark energy and of other theoretically related problems is given for example in \cite{Alexander:2007xx,Alnes:2005rw,GarciaBellido:2008nz,GarciaBellido:2008gd,GarciaBellido:2008yq,February:2009pv,Uzan:2008qp,Quartin:2009xr,Clarkson:2007bc,Zuntz:2011yb,Ishibashi:2005sj,Bolejko:2011ys}.

In this paper we study the anaytical solution of the inversion problem (IP) which consists in matching exactly the terms of the low redshift expansion for the relevant cosmological observables, in particular the role played by central smoothness.
As observed in \cite{Clifton:2008hv} the smoothness of the inhomogeneity profile is important to allow to distinguish these LTB models from $\Lambda CDM$, which is consistent with the conclusion that the solution of the IP is possible for both the luminosity distance $D_L(z)$ and the red-shift spherical mass energy $\rho_(z)$ only if we allow a not smooth radial matter profile \cite{Romano:2009mr}
, while smooth models \cite{Romano:2009ej} can be distinguished from $\Lambda CDM$, at least in principle, without any ambiguity.
In these previous investigations the smoothness condition consisted in assuming that the function necessary to define the LTB geometry did not contain any first order derivative term at the center.
In this paper we show that in order to ensure smoothness at higher orders it is necessary to impose the condition that any odd derivative of the functions defining the model vanishes at the center, and study its implication on the solution of the inversion problem for $D_L(z)$, including for the first time, also the constraint coming from the age of the Universe.

It is important to observe that according to our definition of the IP we look for the LTB models which match the coefficients of the redshift expansion of the observables corresponding to the best fit flat $\Lambda CDM$ models, but since these models depend only on the two independent parameters $H_0,\Omega_M$, LTB models have a higher number of parameters, implying that they could actually provide an even better fit of experimental data.
Our definition of IP is nevertheless quite  natural  from a mathematical point of view since it consists in matching the red-shift space theoretical predictions of different models within the range of validity of the Taylor expansion> The analytical formulae can also be used for low-redshift data fitting, with the advantage of not depending on any functional ansatz.
We can for example draw contour plots for the coefficients of the expansion of the functions defining the LTB model , in the same way we do for $\Omega_{\Lambda}$.
As far as observations are concerned it would in be more important to fit directly actual experimental data rather than mimic the best fit theoretical $\Lambda CDM$ model, so our conclusions should be considered keeping this in mind, and we leave to a future work the analysis of experimental data.

This kind of analysis has been performed by different groups \cite{Biswas:2010xm,Moss:2010jx,Marra:2011ct}, and recently \cite{Riess:2011yx} it has been claimed that accurate measurements of $H_0$ are sufficient to rule out best fit void models. This claim has been corrected in \cite{EneaRomano:2011aa}, and the dimensionless form in which we re-write the equations to solve the IP, makes it evident even in the low-redshift expansion approach we adopt, that $H_0^{LTB}$ can accommodate any observed $H_0^{obs}$.

The paper is organized as follows.
We devote a section to the general requirements for the LTB solution to be smooth, i.e. infinitely differentiable, at the center, generalizing it to the requirement that all the odd derivatives of the functions of the radial coordinates defining the model should vanish at the center; without this smoothness condition even derivatives of the energy density would diverge at the center.
Using a local Taylor expansion we discuss in details the existence, uniqueness and  degeneracy of the solutions of the IP consisting in matching the observations of the luminosity distance, $H_0$ and the age of the Universe. We classify  models according to the central smoothness and the presence of an isotropic but inhomogeneous big bang.
We perform the some classification using only the first order smoothness condition , and the the higher order ones, showing how these latter strongly impact the number of smooth solutions. Higher order smoothness conditions in fact strongly limit the number of possible  solutions respect to the first order condition. One difference is in the case in which the age constrain is not included and the bang function is zero, in which there is a unique solution for $C^1$ models but no solution for the $C^{i}$ case. Another difference is in the case in which the age constraint is not included and the bang function is not zero, in which the solution is undetermined for both $C^1$ and $C^{i}$ models, but the latter ones have much less residual parametric freedom. 
This implies that any other LTB model able to fit luminosity distance data, satisfy the age constraint and fit some other observable is either not mimicking exactly the $\Lambda CDM$ red-shift space observables theoretical predictions or it is not $C^{\infty}$ smooth.

\section{Lemaitre-Tolman-Bondi (LTB) Solution\label{ltb}}
This is one form to write the Lemaitre-Tolman-Bondi \cite{Lemaitre:1933qe,Tolman:1934za,Bondi:1947av}
\begin{eqnarray}
\label{eq1} %
ds^2 = -dt^2  + \frac{\left(R,_{r}\right)^2 dr^2}{1 + 2\,E(r)}+R^2
d\Omega^2 \, ,
\end{eqnarray}
where $R$ is a function of the time coordinate $t$ and the radial
coordinate $r$, $R=R(t,r)$, $E$ is an arbitrary function of $r$ and $R,_{r}=\partial R/\partial r$.

from the Einstein's equations we obtain
\begin{eqnarray}
\label{eq2} \left({\frac{\dot{R}}{R}}\right)^2&=&\frac{2
E(r)}{R^2}+\frac{2M(r)}{R^3} \, , \\
\label{eq3} \rho(t,r)&=&\frac{2 M,_{r}}{R^2 R,_{r}} \, ,
\end{eqnarray}
with $M=M(r)$ being another arbitrary function of $r$ and the dot denoting
the partial derivative with respect to $t$, $\dot{R}=\partial R(t,r)/\partial t$.
It is possible to introduce the variables
\begin{equation}
 A(t,r)=\frac{R(t,r)}{r},\quad k(r)=-\frac{2E(r)}{r^2},\quad
  \rho_0(r)=\frac{6M(r)}{r^3} \, ,
\end{equation}
so that  Eq.\ (\ref{eq1}) and the Einstein equations
(\ref{eq2}) and (\ref{eq3}) are written in a form 
similar to those for FLRW models,
\begin{equation}
\label{eq6} ds^2 =
-dt^2+A^2\left[\left(1+\frac{A,_{r}r}{A}\right)^2
    \frac{dr^2}{1-k(r)r^2}+r^2d\Omega_2^2\right] \, ,
\end{equation}
\begin{eqnarray}
\label{eq7} %
\left(\frac{\dot{A}}{A}\right)^2 &=&
-\frac{k(r)}{A^2}+\frac{\rho_0(r)}{3A^3} \, ,\\
\label{eq:LTB rho 2} %
\rho(t,r) &=& \frac{(\rho_0 r^3)_{, r}}{3 A^2 r^2 (Ar)_{, r}} \, .
\end{eqnarray}
The solution takes now the form 
\begin{eqnarray}
\label{LTB soln2 R} a(\eta,r) &=& \frac{\rho_0(r)}{6k(r)}
     \left[ 1 - \cos \left( \sqrt{k(r)} \, \eta \right) \right] \, ,\\
\label{LTB soln2 t} t(\eta,r) &=& \frac{\rho_0(r)}{6k(r)}
     \left[ \eta -\frac{1}{\sqrt{k(r)}} \sin
     \left(\sqrt{k(r)} \, \eta \right) \right] + t_{b}(r) \, ,
\end{eqnarray}
where $\eta \equiv \tau\, r = \int^t dt'/A(t',r) \,$ and $A(t(\eta,r),r)=a(\eta,r)$ and $t_{b}(r)$ is another arbitrary function of $r$, called the bang function,
which is an integration constant corresponding to the fact that the big-bang initial singularity can happen at different
times at different distances from the center. The inhomogeneity of the location of the singularities can lead to possible causal separation \cite{Romano:2006yc} between 
the central observer and the spatially averaged region for models
 with positive $a_D$.
There is an open debate about the presence of the bang function in LTB models consistent with observations, since the bang function would correspond \cite{Yoo:2008su,Zibin:2008vk} to large decaying modes in the primordial curvature perturbations in disagreement with the cosmic microwave background (CMB)
observations. Other authors \cite{Celerier:2009sv}
 claim instead that the present CMB observations are not incompatible with an isotropic but inhomogeneous bang function.
In this paper we will not deal with the physical implications of an inhomogeneous big bang, but in the section in which we analyze the possible solutions to the IP we will distinguish between the homogeneous and inhomogeneous big bang case in order to provide a general  analysis of all the possible mathematical cases.  
In the rest of paper we will use this last set of equations.
Without loss of generality we can set 
the function $\rho_0(r)$ to be a constant,
 $\rho_0(r)=\rho_0=\mbox{constant}$, which the choice of coordinates in which $M(r)\propto r^3$.


\section {The inversion problem for $D_L(z)$ }
The general low re-shift formula for the  luminosity distance for a central observer is \cite{Romano:2009mr}
\bea
D_L(z)&=&(1+z)^2r(z)a(\eta(z),r(z))=D_1 z+D_2 z^2+D_3 z^3 + . .\\
D_1&=&\frac{1}{H_0},\nonumber\\
D_2&=&\frac{1}{{2 a_0^3 H_0^4 (2 q_0-1)^{5/2}}}\bigg[\sqrt{2 q_0-1} (-a_0^3 H_0^3 (1-2 q_0)^2 (q_0-1)-2 a_0^2 H_0^3 (4
   q_0^3-3 q_0+1) t^b_1-9 k_1 q_0)+\nonumber\\
   &&+6 k_1 q_0 (q_0+1) X\bigg],\nonumber\\
D_3&=&\frac{}{{4 a_0^6
   H_0^7 (2 q_0-1)^{11/2}}}\bigg[-3 (2 q_0-1) q_0 X (4 a_0^2 H_0^3
   k_1 (1-2 q_0)^2 q_0 (2 a_0 q_0+4 q_0 t^b_1+t^b_1)+\nonumber \\
   &&-4 a_0^2 H_0^2
   k_2 (4 q_0^3-3 q_0+1)+k_1^2 (50 q_0^2+7 q_0-7))+(2
   q_0-1)^{3/2} (2 a_0^6 H_0^6 (1-2 q_0)^4 (q_0-1) q_0+\nonumber\\
   &&+8 a_0^5 H_0^6 (1-2
   q_0)^4 q_0^2 t^b_1+4 a_0^3 H_0^3 k_1 (1-2 q_0)^2 q_0 (5 q_0-1)+\nonumber\\
   &&+2a_0^2 H_0^2 (1-2 q_0)^2 (H_0 k_1 (20 q_0^2+q_0-1) t^b_1-9
   k_2 q_0)+2 H_0^5 (a_0-2 a_0 q_0)^4 (H_0 q_0 (4 q_0+1)
   {(t^b_1)}^2+\nonumber\\
   &&-2 (q_0+1) t^b_2)+9 k_1^2 q_0 (11 q_0-4))+18 k_1^2 \sqrt{2
   q_0-1} (4 q_0+1) q_0^3 X^2\bigg]\,,  \nonumber
\eea
where we have introduced 
\bea
a_0=a(\eta_0,0)=\frac{\tan(\frac{\sqrt{k_0}\eta_0}{2})^2 \rho_0}{3 k_0 (\tan(\frac{\sqrt{k_0}\eta_0}{2})^2+1)}, \label{a0}\\
H_0=\frac{3 k_0^{3/2} \left(\tan(\frac{\sqrt{k_0}\eta_0}{2})^2+1\right)}{\tan(\frac{\sqrt{k_0}\eta_0}{2})^3 \rho_0}, \label{H0}\\
q_0=\frac{1}{2} \left(\tan(\frac{\sqrt{k_0}\eta_0}{2})^2+1\right)\label{q0}\,. 
\eea
It is convenient to re-write it in terms of dimensionless parameters according to:
\bea
K_n&=&k_n(a_0 H_0)^{-(n+2)}, \\
T_b^n&=&t_b^n a_0^{-1}(a_0 H_0)^{-n+1}, \\
\eea

to get for example:
\bea
D_2&=&\frac{6 K_1 q_0 (q_0+1) \sqrt{2 q_0-1} X-(2 q_0-1) \left(q_0 (9 K_1-6
   T_b^1+5)+q_0^3 (8 T_b^1+4)-8 q_0^2+2 T_b^1-1\right)}{2 H_0 (2 q_0-1)^3}. \nonumber \\ 
   &&
\eea 
As expected observable quantities such as $D_L(z)$ do not depend on $a_0$. 
Based on the above results it is convenient to schematically write the coefficients of the expansion of $D^{LTB}_L(z)$ in the form:
\bea
D_2^{LTB}&=&D^{LTB}_2(H_0,q_0,K_1,T_b^1) \,, \\
D_3^{LTB}&=&D^{LTB}_3(H_0,q_0,K_1,K_2,T_b^1,T_b^2) \,,\\
&.......& \,, \\
D_i^{LTB}&=&D^{LTB}_i(H_0,q_0,K_1,...,K_{i-1},T_b^1,...,T_b^{i-1}) \,,
\eea
which clearly shows what is the dependence between the coefficients of the expansion of the function defining the LTB model $K(r),T_b(r)$ and those of the red-shift space observable.

The analytical version of the inversion problem up to order $n$ consists in matching the coefficients of the expansion of $D^{\Lambda CDM}_L(z)$ and $D^{LTB}_L(z)$:
\bea
D^{LTB}_i&=&D^{\Lambda CDM}_i \quad,\quad 1\leq i \leq n \,.
\eea

\section{Central smoothness and observer location fine-tuning}
It has been proved using a local Taylor expansion approach that \cite{Romano:2009ej} only not centrally smooth LTB models can mimick the effects of the cosmological constant 
for more than one observable. 
This is independent from the value of the cosmological constant, and is related to the number of free independent parameters on which the functions defining the solutions 
depend on.
Central smoothness of the LTB solution is also important because is related to the sign of the apparent value \cite{Vanderveld:2006rb} of the deceleration parameter $q_0^{app}$. 
For this reason we will later classify the solutions of the inversion problem according to central smoothness, since the space of not smooth solutions is larger than that of 
smooth solutions, and in same cases there exist no smooth solutions, as shown in the next section.

From a general point of view a function of the radial coordinate $f(r)$ is smooth at the center $r=0$ only if all its odd derivatives vanish there. 
This can be shown easily by looking at the the partial derivatives of even order of this type for example:
\bea
\partial^{2n}_x\partial^{2n}_y\partial^{2n}_z f(\sqrt{x^2+y^2+z^2}) \,,
\eea
where $\{x,y,z\}$ are the cartesian coordinates related to $r$ by $r^2=x^2+y^2+z^2$.
Quantities of the type above diverge at the center if $\partial^{2m+1}_r f(r) \neq0$ for $2m+1<2n$.
If for example the first derivative $f'(0)$ is not zero, then the laplacian will diverge.

The general central smoothness conditions for a function of class $C^{i}$ are:

\bea
k_{2m+1}&=&0 \,, \\
t_b^{2m+1}&=&0 \,, \\
2m+1&<&i \,,
\eea
which must be satisfied for all the relevant odd powers coefficients of the central Taylor expansion.
For definiteness we will present here the results of the calculations to the second order in $r$, corresponding to third order in $z$ for $D_L(z)$. We will show later nevertheless that our conclusions about the existence and uniqueness of the solution are independent of the order
at which we truncate the above expansion, and that in general the expansion of the $D_L(z)$ to the i-th order in $z$ requires the expansion to (i-1)-th in $r$ for $k(r)$ and $t_b(r)$.

A possible criticism about the physical relevance of considering central smoothness as a criterion to study LTB models is that the available observations, such as 
Supernovae for example, are at redshift higher than $0.013$, i.e. relatively far from the central observer. This implies that fitting data with centrally smooth models can 
still give good results since there are not sufficiently low redshift observations to affect statistically the data fitting. If we assume nevertheless that the $\Lambda CDM$ 
model is also phenomenologically valid at redshift lower than the range of the currently available cosmological observations, we should expect that the goodness of the fit at 
very low redshift should be quite poor for smooth LTB  models, as predicted by the low redshift expansion analysis. This is the reason why smooth models can give good results in fitting supernovae data, even if their low red-shift luminosity distance is not mimicking exactly the $\Lambda CDM$ model, since most of the data used to compute the goodness of the fitting are at higher red-shift where the central smoothness is not so important anymore. On the other side the existence of a center is a distinctive feature of LTB models and consequently the behavior of the solution around the center cannot be neglected.

 From a mathematical point of view it has been proved
 \cite{Romano:2009ej} that central smoothness is a key characteristic for a LTB model to be able to mimick $\Lambda CDM$ for more than one observable, independently from the
 value of the cosmological constant. From a data fitting point of view, this cannot be checked yet due to the lack of sufficiently low redshift observations, which does not
 diminish the relevance of such an important mathematical feature of LTB models as viable cosmological models. 

Finally we note that the LTB models which have been claimed to be incompatible with recent accurate $H_0$ measurements are based on the assumption of the central location of the observer, which we also make in our analysis.
In these models the observed dipole anisotropy of the  CMB radiation is associated to our velocity relative to the CMB, and is the same as in a $\Lambda CDM$ model, since for a central observer there is no additional anisotropy added by the propagation of photons in a spherically symmetric inhomogeneous space. Alternatively the dipole anisotropy component of the CMB could be considered the effect of an off-center position of the observer in a spherically symmetric space, due to the fact that space is not isotropic around the off-center observer. 
Combining Supernovae observations and the CMB anisotropy dipole it has been found \cite{Alnes:2006uk} that the location of the observer has to be within one percent of the void radius, a fine tuning which is in clear violation of the Copernican principle, and remains the main criticism against these models.   
Independently from the position of the observer LTB solutions have a center, and the smoothness conditions would still apply even to the case of an off-center observer.
While it could be interesting to investigate how  an off-center position of the observer could allow to improve the solution of the inversion problem, it is clear that since the only additional parameter would be the distance from the center,  the general conclusions achieved in this paper will not be substantially affected up to an additional residual degree of freedom in determining the solution.

\section{How many independent parameters determine locally a LTB dust model?}

In order to fit observation with a LTB dust model we need to understand well how many are the really independent free parameters. From the previous section we know that 
that any expression containing $\rho_0,\eta_0,k_0$ can be re-expressed in terms of $a_0,H_0,q_0$, and that introducing appropriate dimensionless quantities $a_0$ does not 
appear explicitly in observables such as $D_L(z)$.
This suggests that there are only two independent parameters to fix in order to fit observations, and we will prove this in a more rigorous way.

We can start from rewriting the equations eq.(\ref{a0}-\ref{q0}) in the following form:

\bea
1=\frac{\Omega_m}{K_0}\frac{2 q_0-1}{q_0}&=&f_1(q_0,K_0,\Omega_m) \,,\\
1=\frac{2 q_0}{\Omega_m}\left(\frac{K_0}{2 q_0-1}\right)^{3/2}&=&f_2(q_0,K_0,\Omega_m)\,,\\
1=\frac{1}{2 q_0}\left[\tan{\left(\frac{\sqrt{K_0}T_0}{2}\right)}^2+1\right]&=&f_3(q_0,K_0,T_0)\,,
\eea
where we have used introduced the dimensionless parameter $\Omega_m$ and $T_0$ according to:

\bea
\rho_0&=&3 a_0^3 H_0^2 \Omega_m \,,\\
\eta_0&=&T_0(a_0 H_0)^{-1}\,.
\eea
Another possible constraint could come from the age of the Universe.
Since the dust models we are considering cannot be used to describe the universe during the radiation dominated stage, they cannot really be considered fully realistic cosmological models.
Nevertheless, since most of time the Universe has been matter dominated, we can safely neglect the contribution to the age of the Universe coming from the radiation dominated era.
Using the exact analytical solution $t(\eta,r)$ we can write:

\bea
T_U =H_0 t(\eta_0,0)=  \frac{\Omega_m}{2 k_0}\left[T_0-\frac{1}{\sqrt{K_0}}\sin{\left(\frac{\sqrt{K_0}T_0}{2}\right)}\right]=f_4(\Omega_m,K_0,T_0) \,,
\eea
where we have introduced the dimensionless quantity $T_U=t_U H_0$, which is the age of the Universe in units of $H_0^{-1}$.

We have four equations four parameters ${K_0,T_0,q_0,\Omega_m}$, while $H_0$ remains free.
This is manifest from the fact that the equations are written in terms of dimensionless parameters which are independent from the value of $H_0$. 

\section{$H_0$ observations and LTB models}
The value of $H_0$ is defined observationally in a model independent way as:
\bea
(H_0^{obs})^{-1}&=&\lim \limits_{z \to 0} \frac{d D_L^{obs}(x)}{d x}_{x=z} \,.
\eea
Let us now expand locally the observed luminosity distance :

\bea
D_L^{obs}(z)&=& D^{obs}_1 z+ D^{obs}_2 z^2+ ..= \frac{1}{H_0^{obs}} z+D^{obs}_2 + ..
\eea
In order to avoid confusion we will denote here all the quantities related to a LTB model with superscripts according to:

\bea
H_0^{LTB}&=&H_0^{LTB}(\rho_0,k_0,\eta_0)=H_0^{LTB}(\Omega_m,K_0,T_0) \,,\\
q_0^{LTB}&=&q_0^{LTB}(K_0,T_0) \,.
\eea
We will also limit ourselves to the homogeneous bang function case, i.e. set $t_b=0$, and mention later how including the bang function will make our arguments even stronger.
According to the expansion for the luminosity distance obtained previously we have

\bea
D_1^{LTB}&=&\frac{1}{H_0^{LTB}} \,,\\
H_0^{LTB} D_2^{LTB}&=&f_5(q_0^{LTB},K_1,T^b_1)\,,\\
H_0^{LTB} D_3^{LTB}&=&f_6(q_0^{LTB},K_1,K_2,T^b_1,T^b_2)\,, \\
&.........& \,,\\
H_0^{LTB} D_i^{LTB}&=&f_{i+3}(q_0^{LTB},K_{1...i-1},T^b_{1..i-1}) \,. \nonumber 
\eea
The advantage to write in this form the coefficients of the expansion for the luminosity distance is that in this way we can clearly understand the role of $H_0$ as the natural scale of the problem, which is set by $H_0^{obs}$.

Contrary to some previous attempts \cite{Alexander:2007xx} to fit observation with void models the correct way to interpret this equations is not to look for the particular set of parameters which give $H_0^{LTB}=H_0^{obs}$, but to directly fix $H_0^{LTB}=H_0^{obs}$ and then look for the dimensionless parameters which solve the inversion problem, i.e. specify the LTB model in agreement with observations.
This means that there is always a set of LTB solutions in agreement with any value of $H_0^{obs}$, as long as low redshift observations for the luminosity distance are concerned, and an arbitrarily precise measurement of $H_0^{obs}$ is not by itself able to rule out void models\cite{Riess:2011yx}, but only able to establish accurately the natural scale in terms of which to define the model.

In order to design a model in good agreement at low redshift with observations we can match the Taylor expansion of the observed luminosity distance with the corresponding LTB terms :

\bea
D_i^{obs}=D_i^{LTB} \,.
\eea
Combining the above equations with the ones derived in the previous section we finally get:
\bea
H_0^{obs}&=&H_0^{LTB} \,,\\
1&=&f_1(q_0^{LTB},K_0,\Omega_m) \,,\label{f1} \\
1&=&f_2(q_0^{LTB},K_0,\Omega_m) \,,\label{f2}\\
1&=&f_3(q_0^{LTB},K_0,T_0) \,,\label{f3}\\
T_U&=&f_4(\Omega_m,K_0,T_0) \,,\label{f4}\\
H_0^{obs} D_2^{obs}&=&f_5(q_0^{LTB},K_1,T^b_1)\,, \label{f5}\\
H_0^{obs} D_3^{obs}&=&f_6(q_0^{LTB},K_1,K_2,T^b_1,T^b_2)\,, \label{f6}\\
&.........& \,,\\
H_0^{obs} D_i^{obs}&=&f_{i+3}(q_0^{LTB},K_{1...i-1},T^b_{1..i-1}) \,.\label{fi}\nonumber 
\eea
As mentioned previously the advantage of writing the equations in this form is to clearly identify the role of $H_0^{LTB}$ as  the scale in terms of which to define the LTB model, which is given directly from the first equation. 
We will analyze in the rest of the paper the different possible cases according to central smoothness and the presence of a bang function $t_b(r)$.
One last important point is that $H_0^{LTB}$ is a purely local quantity, i.e. only depends on the central value of $k(r),\eta$ and  $\Omega_m$, implying that the Taylor expansion approach we adopted gives exact results as long as $H_0$ is concerned.

\section{Inversion problem solutions for $C^{1}$ models}
The number of equations to be satisfied to ensure the matching of the luminosity distance up to
i-th order is $4+(i-1)$, where we have $(i-1)$ constraints because we consider the matching of the first term $D_1^{LTB}$equivalent to fixing the scale in terms of which to define the rest of the parameters of the model. 

\vspace{1cm}

\begin{table}[htbp]
		\begin{tabular}{|l|l|l|l|l|} 
		\hline
		\multicolumn{5}{|c|}{Age constraint}\\ 
		\hline
	  LTB type & Not $C^1$, $t_b\neq0$ & Not $C^1$, $t_b=0$ & $C^1$, $t_b\neq0$ & $C^1$, $t_b=0$ \\ \cline{1-5} 
		Free parameters &4+2(i-1) & 4+(i-1) & 4+2(i-1)-2 & 4+(i-1)-1 \\ \hline
		Constraints & 3+(i-1)+1 & 3+(i-1)+1& 3+(i-1)+1& 3+(i-1)+1 \\ \hline
		Solution & Undetermined & Unique & No& No  \\ \hline
		\end{tabular}
	\caption{The number of parameters on which LTB models depend expanding the luminosity distance to i-th order are shown according to the central behavior of the energy density (smooth/not smooth) and the presence of a homogeneous bang function $t_b(r)$. In this case we are including the constraint coming from the age of the Universe.}
	\label{tab:I}
\end{table}
\begin{table}[htbp]
		\begin{tabular}{|l|l|l|l|l|} 
		\hline
		\multicolumn{5}{|c|}{No Age constraint}\\ 
		\hline
		LTB type & Not $C^1$, $t_b\neq0$ & Not $C^1$, $t_b=0$ & $C^1$, $t_b\neq0$ & $C^1$, $t_b=0$ \\ \cline{1-5} 
		Free parameters &4+2(i-1) & 4+(i-1) & 4+2(i-1)-2 & 4+(i-1)-1 \\ \hline
		Constraints & 3+(i-1) & 3+(i-1)& 3+(i-1)& 3+(i-1) \\ \hline
		Solution & Undetermined & Undetermined & Undetermined & Unique \\ \hline
		\end{tabular}
	\caption{The number of parameters on which LTB models depend expanding the luminosity distance to i-th order are shown according to the central behavior of the energy density (smooth/not smooth) and the presence of a homogeneous bang function $t_b(r)$. In this case we are not including the constrain coming from the age of the Universe.}
	\label{tab:I}
\end{table}

In order to understand the content of the tables it can be seen from eq.(\ref{f1}-\ref{f3}) that independently of the order of the matching of the Taylor expansion there always are four free parameters $q_0,K_0,T_0,\Omega_m$ present and three constraints. When including the age of the universe  then we have an extra constraint. From the matching of the coefficients up to i-th order of the luminosity distance expansion instead we get $2(i-1)$ parameters corresponding to $K_{i-1},T_{i-1}$, coming from $D_{i}^{LTB}$. Imposing the $C^1$ condition or setting to zero the bang function $t_b(r)$ reduces the $2(i-1)$ parameters to the values reported in the tables. 

The condition of the existence of a solution of the inversion problem is:
\bea
\# parameters \geq \# constraints \quad,\quad i\geq 2
\eea
In the case of equality the solution is unique.

\subsection{Analysis of the relevant cases}

In the following it should be noted, as mentioned above, that we are assuming by construction that $H_0^{LTB}=H^{obs}$(which is always possible), and that all the other parameters defining a LTB are dimensionless.
The existence of a solution of the inversion problem can be easily determined by looking at the number of free parameters and the number of unknowns reported in the tables but there are some cases in which some extra explanations are required.
\subsubsection{$C^{1}$ with age constraint}

When the functions defining the model are $C^1$ at the center, i.e. when $K_1=T^b_1=0$,  we have no solution if we impose the constraint coming from the age of the Universe, with or without the bang function $t_b(r)$.
Apparently for $i\geq3$ we have more free parameters than constraints, but for $i=2$ we have that the system of five equations (\ref{f1}-\ref{f5}) has only four unknowns and the higher order parameters $K_{i-1},T^b_{i-1}$ appearing in $D_{i}^{LTB}$ do not have any effect on equations (\ref{f1}-\ref{f5}).
If we do not include the constrain from the age of the Universe in eq.(\ref{f4}), the solution  is actually undetermined, because in this case the system of four equations (\ref{f1}-\ref{f3},\ref{f5}) has a unique solutions, but higher order terms will introduce extra degeneracy, since every term $D_i$ will contain two independent parameters $K_{i-1},T^b_{i-1}$, and only one more constraint equation.

\subsubsection{Not $C^{1}$ without bang function and with age constraint} 

The solution is unique in this case, since the number of parameters is equal to number of constraints at any order.
This implies that any other LTB which which can fit observational data is either not mimicking exactly the the cosmological constant or, if it does, it is not smooth. 
This is due to fact that we are imposing very restrictive constraints on the functions defining LTB models, while data fitting allows for much more freedom due to the tolerance introduced in the statistical analysis, which can allow good data fitting even if the low-redshift observable are not exactly the same, and also because there are very few data at very low-redshift, making this mismatch not statistically important in evaluating the goodness of the fit.

\subsubsection{Not $C^{1}$ case with  bang function} 
When we allow for not $C^1$ models with bang function, we gain some local extra parametric freedom , which implies that the age of the Universe and luminosity distance are not enough to uniquely determine a LTB model in agreement with observations, leaving more freedom to fit other observables.

\section{Inversion problem solutions for $C^{i}$ models}
In this section we will consider the implication of the $C^{i}$ conditions on the solution of the inversion problem.
In general the space of free parameters is reduced further respect to the $C^1$ case, making the solution of the inversion problem some time impossible.
As already observed for the $C^{1}$ models it can be seen from eq.(\ref{f1}-\ref{fi}) that there are four free parameters $q_0,K_0,T_0,\Omega_m$ present independently of the order of the matching of the Taylor expansion, and  $2(i-1)$ parameters corresponding to $K_{i-1},T_{i-1}$, coming from $D_{i}^{LTB}$. Imposing central smoothness or setting to zero the bang function $t_b(r)$ reduces the $2(i-1)$ parameters to the values reported in the tables.
The number of constraints are the same as the ones in the previous sections.
\begin{table}[htbp]
		\begin{tabular}{|l|l|l|l|l|} 
		\hline
		\multicolumn{5}{|c|}{Not smooth}\\ 
		\hline
		Age constraint & \multicolumn{2}{c|}{Yes} & \multicolumn{2}{c|}{No} \\ 
		\hline
		LTB type & $t_b\neq0$ & $t_b=0$ &  $t_b\neq0$ & $t_b=0$ \\ \cline{1-5} 
		Free parameters &4+2(i-1) & 4+(i-1) & 4+2(i-1) & 4+(i-1)  \\ \hline
		Constraints & 3+(i-1)+1 & 3+(i-1)+1 & 3+(i-1)& 3+(i-1) \\ \hline
		Solution & Undetermined & Unique & Undetermined & Undetermined  \\ \hline
		\end{tabular}
	\caption{The number of parameters on which LTB models depend expanding the luminosity distance to i-th order are shown in the not centrally smooth case according to the presence of a homogeneous bang function $t_b(r)$ and the age constraint.}
	\label{tab:I}
\end{table}

\begin{table}[htbp]
		\begin{tabular}{|l|l|l|l|l|}  
		\hline
		\multicolumn{5}{|c|}{Smooth without age constraint}\\ 
		\hline
		Matching to i-th order in $D_L(z)$ & \multicolumn{2}{c|}{Even} & \multicolumn{2}{c|}{Odd} \\ \hline
		LTB type & Smooth, $t_b\neq0$ & Smooth, $t_b=0$ &Smooth, $t_b\neq0$ &Smooth, $t_b=0$ \\ \cline{1-5} 
		Free parameters &4+(i-2) & 4+(i-2)/2 & 4+(i-1) & 4+(i-1)/2 \\ \hline
		Constraints & 3+(i-1) & 3+(i-1)& 3+(i-1)& 3+(i-1) \\ \hline
		Solution & Unique & NO & Undetermined & NO \\ \hline
		\end{tabular}
	\caption{The number of parameters on which LTB models depend expanding the luminosity distance to i-th order are shown in the smooth case without age constraint and according to the presence of a homogeneous bang function $t_b(r)$.}
	\label{tab:II}
\end{table}

\begin{table}[htbp]
		\begin{tabular}{|l|l|l|l|l|}  
		\hline
		\multicolumn{5}{|c|}{Smooth with age constraint}\\ 
		\hline
		Matching to i-th order in $D_L(z)$ & \multicolumn{2}{c|}{Even} & \multicolumn{2}{c|}{Odd} \\ \hline
		LTB type & Smooth, $t_b\neq0$ & Smooth, $t_b=0$ &Smooth, $t_b\neq0$ &Smooth, $t_b=0$ \\ \cline{1-5} 
		Free parameters &4+(i-2) & 4+(i-2)/2 & 4+(i-1) & 4+(i-1)/2 \\ \hline
		Constraints & 3+(i-1)+1 & 3+(i-1)+1& 3+(i-1)+1& 3+(i-1)+1 \\ \hline
		Solution & NO & NO & NO & NO \\ \hline
		\end{tabular}
\caption{The number of parameters on which LTB models depend expanding the luminosity distance to i-th order are shown in the smooth case with age constraint and according to the presence of a homogeneous bang function $t_b(r)$.}
	\label{tab:III}
\end{table}		
\subsubsection{Analysis of the relevant cases}
The information about the existence of a solution of the inversion problem can be obtained by comparing the number of the free parameters to the number of constraints, but some cases require some further analysis.
\subsubsection{$C^{i}$ models}

The distinction between odd and even order of the Taylor expansion for the $C^{i}$ case comes from the fact that if $i$ is odd, then we need to expand to $(i-1)$-th order in $r$, which is an even number, can be included in the expansion, and so we have some new free parameter. On the contrary if $i$ is even we should expand to the $(i-1)$-th in $r$, which  is odd, and because of the smoothness conditions the $(i-1)$th order cannot be included, but we still have an additional constraint coming from the $D^{i}_L(z)$ matching condition, making the inversion problem more difficult to solve. Intuitively this can be interpreted as the fact that smooth models have less degrees of freedom to satisfy the necessary observational constraints, and going to higher order in the redshift expansion does not always add new free parameters because of the smoothness conditions which require all the odd derivatives for the the functions $k(r)$ and $t_b(r)$ to vanish at the center. 
These are the general properties of the solution of the inversion problem for $D_L(z)$ for $C^{i}$ models, and the differences respect to the $C^{1}$ case:
\begin{itemize}
\item If we include the age constraint there is no solution. This is due to the same reason why there is no solution also in the $C^{1}$ case studied in the previous section, i.e. $D_2$ cannot be matched.
\item If we do not include the age constraint and $t_b(r) \neq0 $ there is a slightly undetermined solution when $i$ is odd  , but only one parameter is left free, the one which is not fixed by the age constraint. This residual freedom is less than in $C^{1}$ model, where it increases as with the matching order $i$.
In the case the matching order is even the solution is unique.
\item  If the age constraint is not included and $t_b(r)=0$ there is no solution, while in the $C^{1}$ case there was a unique solution.
\end{itemize}

\section{Conclusion}
We have extended previous studies about central smoothness in LTB models, by considering the implications of higher order smoothness conditions.
Using a low redshift expansion we have analyzed in detail what are the the constraints that a LTB model has to satisfy in order to be in agreement with $H_0$, luminosity distance and the age of the Universe observations.
The constraint on the age of Universe is actually exact, since the analytical solution can be used to impose it.
We have also given a systematic analysis of the existence of solutions to the inversion problem in the different cases depending on the central smoothness and the presence of a bang function $t_b(r)$, comparing the $C^{1}$ solutions investigated in previous studies to the more general $C^{i}$ case.
For both $C^{1}$ and $C^{i}$ models there is not solution because of the impossibility to match the second order $D_2$term in the expansion of $D_L(z)$. One difference is in the case in which the age constrain is not included and the bang function is zero, in which there is a unique solution for $C^1$ models but no solution for the $C^{i}$ case.
Another difference is in the case in which the age constrain is not included with non zero bang function, in which the solution is undetermined for both $C^1$ and $C^{i}$ models, but the latter ones have much less residual parametric freedom.
Our formulation of the inversion problem uses dimensionless parameters to define the LTB model, which make evident the freedom of fitting any $H_0$ observed value, correcting some previous claims that this last observable would be enough to rule out these models.
The dimensionless formulation of the inversion problem we have derived has the advantage to express the LTB solution directly in terms of observable quantities, which is more transparent than deducing $H_0$ from other parameters of the void model as done in \cite{Biswas:2010xm} for example. 
These results can be used to define a general algorithm to look for LTB models in agreement with both low and high redshift observations, using $H_0^{obs}$ as the fundamental scale in terms of which to define all the other parameters and functions determining the model.
Our results imply that any  LTB model able to fit luminosity distance data, satisfy the age constraint and fit some other observable is either not mimicking exactly the $\Lambda CDM$ red-shift space observables theoretical predictions or it is not $C^{\infty}$ smooth.

While the present paper was focused on pressureless LTB models, the same arguments could be applied to LTB models in presence of a cosmological constant $\lambda$ as proposed in  \cite{Romano:2010nc,Romano:2011mx,Romano:2012gk,Romano:2012kj}, and in this case even small inhomogeneities could have an important effect on the apparent value $\lambda$ and on the equation of state of dark energy.
We will consider this important effects in separate works, underlying nevertheless their general importance since they do not require large voids and should be taken properly into account even in standard $\Lambda CDM$ models.

\begin{acknowledgments}
I thank the members of the Dark energy LeCosPa working group, Misao Sasaki and Marco Regis for comments and discussions.  I also thank the CERN theoretical division for its support and hospitality.
\end{acknowledgments}

\end{document}